\documentclass[prl, reprint]{revtex4-2}
\usepackage{graphicx}
\usepackage{float}
\usepackage{subcaption}
\usepackage{dcolumn}
\usepackage{bm}
\usepackage{amsmath}
\usepackage{soul}
\usepackage{xcolor}
\usepackage{hyperref}

\makeatletter
\long\def\@makecaption#1#2{%
  \par
  \begingroup
    \small\rmfamily
    \begingroup
      \samepage
      \flushing
      \let\footnote\@footnotemark@gobble
      \@make@capt@title{#1}{#2}\par
    \endgroup
  \endgroup
  \vskip\belowcaptionskip
}
\makeatother

\begin{document}


\title{Self-arresting vortex rearrangements in pinned spinning-down superfluids}
\author{S V Anantharaman}\email{anantharaman.sv\_phd20@ashoka.edu.in\\ananth.academic@gmail.com}
\author{Dipankar Bhattacharya}\email{dipankar.bhattacharya@ashoka.edu.in}
\affiliation{%
 Department of Physics, Ashoka University, Sonipat, 131029, India
}%

\date{\today}

\begin{abstract}
Superfluids, in a spinning-down container with pinning centres, exhibit sudden collective migration of quantized vortices. The onset, progress, and termination of this process remain poorly understood. Here, we show that: 1) the first exodus of vortices occurs when a sector of the vortex array, having a minimum opening angle, is vacated close to the wall; 2) driven by this exodus, the superfluid flow near the boundary continually decreases in strength and arrests the exit; and 3) the resultant depletion region has well-defined dimensions---governed primarily by the critical unpinning velocity, the container size, and the initial rotation rate---that limit the angular momentum transferred to the container. We conclude by discussing the relevance to pulsar glitches.
\end{abstract}

\maketitle


\textit{Overview.--} Superfluids and quantum condensates rotate by forming vortices having quantized circulation, arranged in an Abrikosov lattice. The macroscopic velocity field of such a fluid mimics that of a solid body rotating in synchrony with the container \cite{onsager1949, feynman1955}. The fluid remains in-step with a decelerating vessel by reducing the number density of vortices in proportion to the rotation rate, with the outermost vortices annihilating at the container wall. Throughout this process, the array remains in an Abrikosov configuration, only with a continually diminishing vortex density. As a result, the recorded spin-down of the container is smooth. This behaviour is drastically altered by pinning centres, non-superfluid features co-rotating with the container, to which vortices attach. In the presence of vortex pinning, the spin-down of a decelerating container is punctuated by sudden spin-ups \cite{tsakadze1980, skrbek2026}. Pulsar glitches are thought to be astrophysical manifestations of this phenomenon \cite{anderson1975, sauls1989, graber2017, antonopoulou2022}. Several numerical studies, of microscopic (quantum) and mesoscopic (point-vortex) approaches, demonstrate the underlying cause of the spin-up to be a sudden rearrangement of the array during which several vortices unpin and exit the superfluid \cite{warszawski2011, howitt2020, poli2023, liu2024, anantharaman2026}. A recent macroscopic (hydrodynamic) exploration suggests a similar behaviour of the coarse-grained vorticity \cite{perkins2026}. However, the conditions that determine the course of such a rearrangement remain under-examined. In this letter, we outline the mechanisms at play, formulate the conditions relevant to the first spin-up, and demonstrate the existence of characteristic azimuthal and radial scales defining the resulting vortex-scarce region, governed mainly by three quantities: the critical local flow required to unpin a vortex from a pinning site ($v_{\text{crit}}$), the radius of the container ($R$), and the initial rotation rate of the superfluid ($\Omega_0$). The spatial extent of the vacancy places a bound on the net angular momentum transferred to the container, prompting us to estimate the characteristic scales and the corresponding glitch sizes for the Crab and the Vela pulsars.\looseness=-1

\textit{Pinned spinning-down superfluid.--} Consider a superfluid initially in synchrony with its cylindrical container (having radius $R$ and undergoing counter-clockwise rotation at a rate $\Omega_{0}$) and assume that the vortices constituting the array remain rectilinear at all times. This assumption allows us to capture the dynamics by studying only a cross-section of the fluid, but precludes the ability to treat turbulence \cite{donnelly1991}. The vortex density of such a uniform array, $n_{v}$, is related to the rotation rate through the Onsager-Feynman relation, $n_v\kappa = 2\Omega_{0}$, where $\kappa = h / m$ is the quantum of circulation, with $h$ being Planck's constant and $m$ being the mass of the relevant bosonic constituent. The macroscopic velocity field due to this configuration, in the lab frame and described in polar coordinates, is given by $\mathbf{v_{s}}(r,\theta) = r\Omega_{0}~\mathbf{\hat{\theta}}$, where $r$ is the radial distance from the centre of the container to the superfluid element under consideration and $\theta$ is the angle struck with respect to the positive X-axis. Further, assume that the pinning sites are area-filling and uniformly distributed, lending every pinned vortex a well-defined location-independent force, or equivalently a critical unpinning speed $v_{\text{crit}}$. A vortex unpins when the local flow speed in the container's frame exceeds $v_{\text{crit}}$, after which it simply follows the flow, getting recaptured only upon entering a sub-critical region. When the container is subjected to a constant spin-down, a rotational lag develops between the superfluid and the pinned vortex array, growing linearly with time. The velocity lag, during this phase, is largest at the boundary. In due course, the lag at the boundary reaches the unpinning threshold, when the velocity field of the superfluid, in the container's frame, is given by $\mathbf{v_{s}} = v_{\text{crit}}(r/R)~\mathbf{\hat{\theta}}$. Here on, flow fields are treated in the container's frame of reference, unless stated otherwise.

\begin{figure}[b]
\vspace{-2pt}
    \centering
    \includegraphics[width=\linewidth]{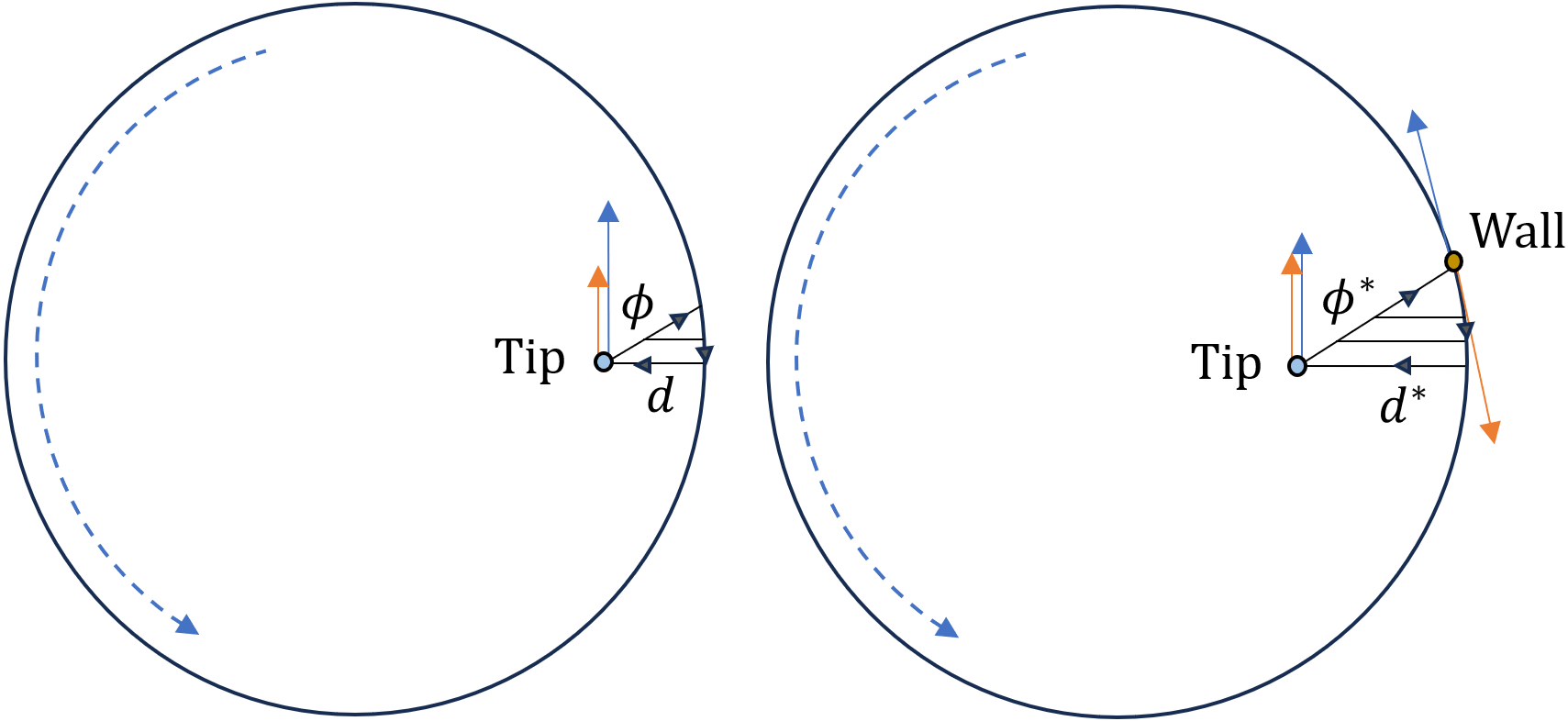}
    \caption{Schematic of the superfluid flow in the presence of a sector devoid of vortices. Striped region indicates a vortex-free zone, with the rest of the container having a uniform distribution of vortices. Blue dashed arrow conveys the container's direction of rotation. Blue and orange solid arrows specify the initial tangential flow and the effect of the sector, respectively. Purple arrowheads indicate the direction of flow on the contours of the vacated region due to the missing vortices. Left) Perturbation - A small sector near the boundary, of depth $d$ and opening angle $\phi$, is depleted of vortices, increasing the tangential flow at the tip. Right) Recapture - The sector is much larger than it was initially, now described by a depth $d^{*}$ and an opening angle $\phi^{*}$. The net flow at the tip is above critical, whereas that at the wall (where the forward arm of the sector meets the boundary) is exactly zero.}
\label{fig:1}
\vspace{-4pt} 
\end{figure}

\textit{Perturbation.--} In an infinitely large cylindrical container, all the vortices at the boundary experience the same local flow. However, in a finite-sized setup, the triangular lattice of the vortex array, being incommensurate with the boundary, develops imperfections near the container wall, leading to the breaking of cylindrical symmetry. We assume that this, along with other possible defects, biases the initial unpinning to occur in a localized region near the periphery, resulting in a vacated sector having a small depth $d$ and an opening angle $\phi$. Notice that the removal of an arbitrary vortex from an initially uniform array leads to an increase in the tangential flow at points radially inward of the vacancy, and a decrease at radially outward points. For a depleted sector, this translates to an augmented tangential field at points radially inward from the centroid, with the tip of the sector experiencing maximum increase, as illustrated in Fig. \ref{fig:1}. The vacancy also affects the radial component, making it non-zero. But, for a small sector, this contribution diminishes close to the container wall, where the radial velocity of the superfluid approaches zero. We impose this boundary condition by introducing image vortices outside the container. Integrating the contributions from all missing vortices, the altered field at the tip, to linear order in $d$, and for small $\phi$, is found to be
\begin{equation}
    \mathbf{v_{tip}} = \left[v_{\text{crit}}\frac{R-d}{R}+\frac{\Omega_{0}d}{\pi}\phi-\frac{\Omega_{0}d}{3\pi} \phi \right] \mathbf{\hat{\theta}},
\label{eq:1}
\end{equation}
where the first term on the right is the lag developed during spin-down, the second captures the effect of the vacated sector, and the third accounts for the boundary by considering the image of the sector. For arbitrarily small values of $\phi$, the flow speed at the tip remains below the critical unpinning threshold. Hence, the first perturbation to the array does not immediately lead to the unpinning of deeper vortices. See End Matter for derivations of Eq.~(\ref{eq:1}) and Eq.~(\ref{eq:3}).

\textit{Exodus.--} As the spin-down induced individual unpinning widens the sector, the flow at the tip becomes critical for a suitably large opening angle $\phi^{*}$ given by
\begin{equation}
    \phi^{*} = \frac{3\pi v_{\text{crit}}}{2R \Omega_{0}},
\label{eq:2}
\end{equation}
leading to the unpinning of a vortex at the tip. The subsequent motion is apparent if we consider the depleted sector as a patch of counter-rotating vortices (having clockwise circulation) superimposed exactly on top of the uniform array (having counter-clockwise circulation), and the tip as having a small but finite curvature. By Stokes' theorem, the vacancy induces a clockwise field along its contours, aligning parallel to the spin-down lag at the tip, and anti-parallel at the wall. See Fig.~\ref{fig:1}. This suggests that a vortex unpinned from the tip would flow along the tangentially forward arm of the sector. A small radially outward displacement of an unpinned vortex along the contour, however, has a two-fold effect, different from the case of removal. It increases the tangential field at radially inward locations (in its wake), while also increasing the field immediately ahead of it, thus promoting the unpinning of vortices in both directions along the sector arm. The exact mechanics of this process invites detailed study. However, we may approximate the exodus along the contour as the continuous loss of thin strips of vortices, such that the depth of the sector proportionally increases, while maintaining the opening angle $\phi^{*}$. Eq.~(\ref{eq:1}) indicates that the flow at the tip, in such a scenario, is always critical, implying continuous unpinning from the tip and a growth of the sector depth.

\textit{Recapture.--} The validity of Eq.~(\ref{eq:1}) in describing the dynamics is crucially dependent on the assumption that all unpinned vortices exit the fluid. Near the wall, however, the radial velocity field is close to zero, raising the question of how a vortex reaches the boundary and annihilates. Dissipation becomes essential in resolving this. Mutual friction imparts a radially outward Magnus force to a vortex moving with the local tangential flow (along the positive $\mathbf{\hat{\theta}}$ direction), resulting in an outward drift and eventual annihilation. However, as highlighted earlier, the growing sector leads to a decreasing tangential field near the container wall, in turn reducing the drift. The field at the boundary $\mathbf{v_{wall}}$, having a non-linear dependence on both the opening angle and the depth, is well approximated, for small $\phi^{*}$, by
\begin{equation}
\begin{aligned}
    \mathbf{{v}_{wall}} = \Biggl[&v_{\text{crit}}- \frac{\Omega_{0} d}{\pi}\phi^{*} \Biggl(\frac{d}{R}\phi^{*}\left(\frac{\pi}{2}-\phi^{*}\right)-\ln{\left(\phi^{*}\right)}\Biggr)\\
    &-\frac{\Omega_{0} d}{\pi}\phi^{*} \left(\frac{3(1-\phi^{*^{2}})}{4}\right)\Biggr]~\mathbf{\hat{\theta}},
\end{aligned}
\label{eq:3}
\end{equation}
where the first term on the right is the lag developed during spin-down, the second captures the effect of the vacated sector, and the third accounts for the boundary by considering the image of the sector. Beyond a certain depth of the vacancy, the tangential field at the boundary aligns with the negative $\mathbf{\hat{\theta}}$ direction. As a result, the front of the exodus in this region experiences a radially inward Magnus drift, effectively halting the exit of vortices. The sector depth $d^{*}$ for which this behaviour emerges is that for which the tangential field at the boundary, given by Eq.~(\ref{eq:3}), vanishes, promoting the recapture of vortices.

\textit{Termination.--} At this juncture, the tip of the sector, being far from the boundary, experiences a field given by
\begin{equation}
\begin{aligned}
\mathbf{v_{tip}} &= \left[v_{\text{crit}}\frac{R-d^{*}}{R}+\frac{\Omega_{0}d^{*}}{\pi}\phi^{*}\right] \mathbf{\hat{\theta}}\\
&= \left[v_{\text{crit}}\left( 1 + \frac{d^{*}}{2R} \right)\right]\mathbf{\hat{\theta}},
\end{aligned}
\label{eq:4}
\end{equation}
where the flow is above critical, indicating further unpinning. However, the ensuing repopulation of vortices near the container wall reduces the flow at the sector tip, contributing to the region becoming sub-critical. Eq. (\ref{eq:1}) suggests that the unpinning completely halts, far from the wall, when the reconfigured region subtends an angle of $\sim\pi v_{\text{crit}}/R\Omega_{0}$. While we do not examine the dynamics of the array during this stage, we note that the shielding effect provided by the re-pinned vortices restricts further rearrangements to occur within the superfluid, allowing no further exit.

\textit{Angular momentum transfer.--} The net angular momentum of a superfluid rotating in synchrony with a cylindrical container is $L_{s} = I_{s} \Omega_{0}$, where $I_s$ is the fluid's moment of inertia about the centre of the container. The contribution of each vortex to the total angular momentum is made explicit through \cite{fetter1965}
\begin{equation}
    L_s = I_s \Omega_{0}\sum_{i=1}^{N}\frac{(R^{2} - r_{i}^{2})}{NR^2/2},
\label{eq:5}
\end{equation}
such that $L_s$ is exactly $I_{s} \Omega_{0}$ for a uniform array. This expression captures the fact that a vortex at an arbitrary radius contributes net angular momentum only to the fluid outward from its location. For a radially outward rearrangement of vortices, the superfluid's net angular momentum continually decreases. By integrating Eq.~(\ref{eq:5}) over the depleted region, we find the reduction in the angular momentum due to a sector of missing vortices, having opening angle $\phi^{*}$ and depth $d$, to be
\begin{equation}
    \Delta L_{s} = I_s\Omega_{0} \left(\frac{d^2}{R^2}\right)  \left(\frac{\phi^{*}}{\pi/2}\left(1-\frac{d}{2R}\right)^{2}\right),
    \label{eq:6}
\end{equation}
where the first bracketed term reflects the radial variation in contribution suggested by Eq.~(\ref{eq:5}) and the second accounts for the sector geometry. Conservation of angular momentum guarantees a proportional increase in the container's rotation rate by an amount $\Delta \Omega_{c} = \Delta L_{s}/I_{c}$, where $I_{c}$ is the moment of inertia of the container about the axis of rotation. The resultant decrease in the velocity lag across the superfluid is then given by
\begin{equation}
    \Delta v_{c} = \frac{\Omega_{0}d}{\pi}\phi^{*}\left[\frac{2I_s}{I_c} \frac{d(R-d)}{R^2} \left(1-\frac{d}{2R}\right)^{2}\right].
    \label{eq:7}
\end{equation}
Eq.~(\ref{eq:1}) and Eq.~(\ref{eq:3}) suggest that the missing sector alters the flow at both the tip and the wall by $\sim \Omega_{0}\phi^{*}d/\pi$.\\For a massive container, such that $I_{s}/I_{c} \leq 1$, Eq.~(\ref{eq:7}) indicates that the transfer of angular momentum plays only a secondary role in the exodus and recapture of vortices. This allows us to characterise the resultant depletion zone by an opening angle $\phi^{*}$ and a sector depth $d^{*}$. The total observed spin-up of the container due to the rearrangement is then well approximated by
\begin{equation}
    \Delta \Omega_{c}^{*} = \Omega_{0}\left[\frac{I_s}{I_c} \frac{{d^{*}}^2}{R^2}  \frac{\phi^{*}}{\pi/2}\left(1-\frac{d^{*}}{2R}\right)^{2}\right].
    \label{eq:8}
\end{equation}
Here, we have not explicitly accounted for factors that alter the rate at which the dynamics unfold, such as the dissipation acting on moving vortices and the spin-down rate of the container. However, assuming such details to be in a range conducive for vortex rearrangement, we find that the emergent spatial scales and the resulting spin-up of the container depend primarily on three quantities: the critical unpinning speed $v_{\text{crit}}$, the radius of the container $R$, and the initial rate of rotation $\Omega_{0}$.\\

\begin{figure}[hb]
    \vspace{18pt}
    \centering
    \includegraphics[width=0.4\linewidth]{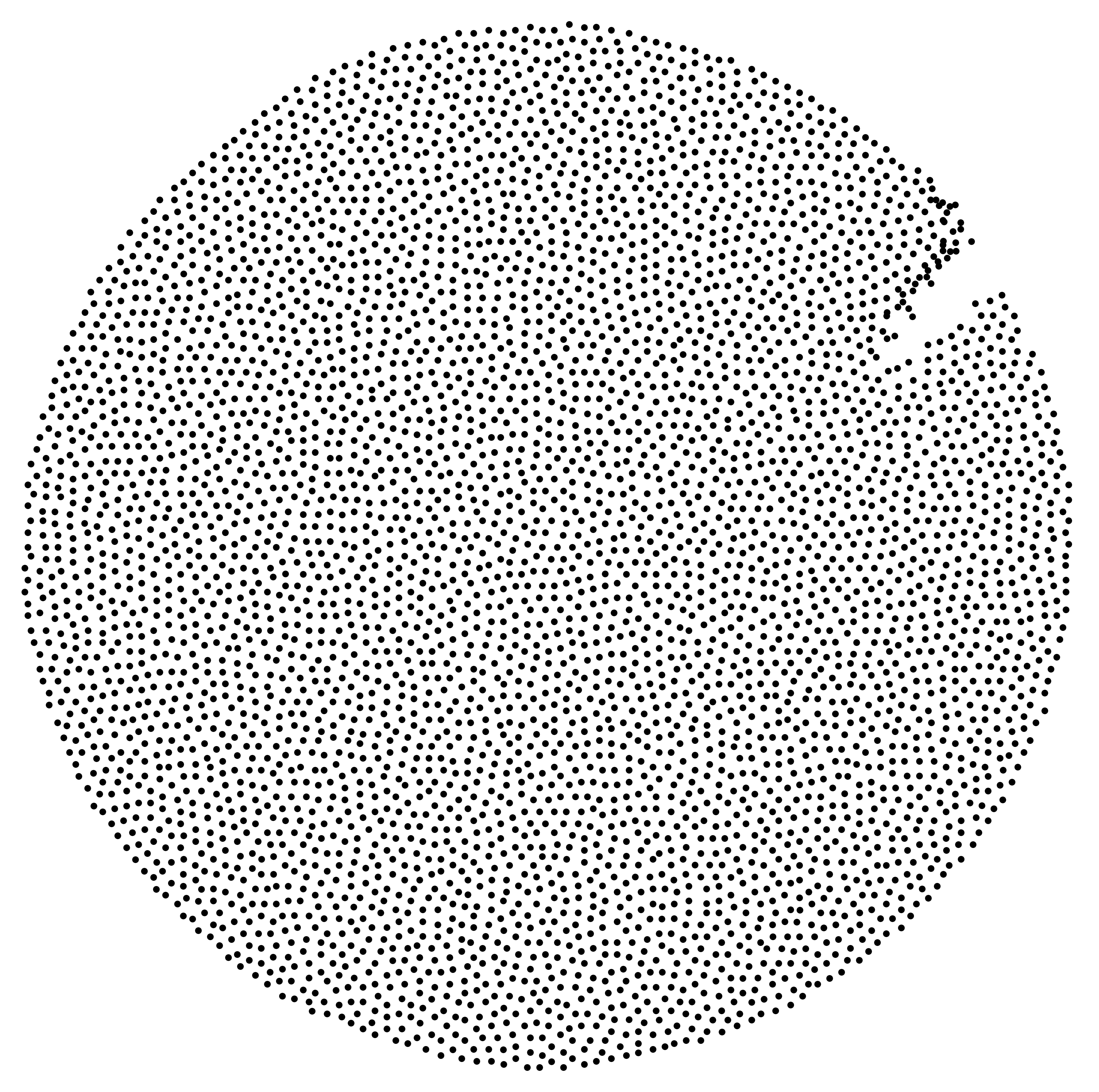}
    \includegraphics[width=0.4\linewidth]{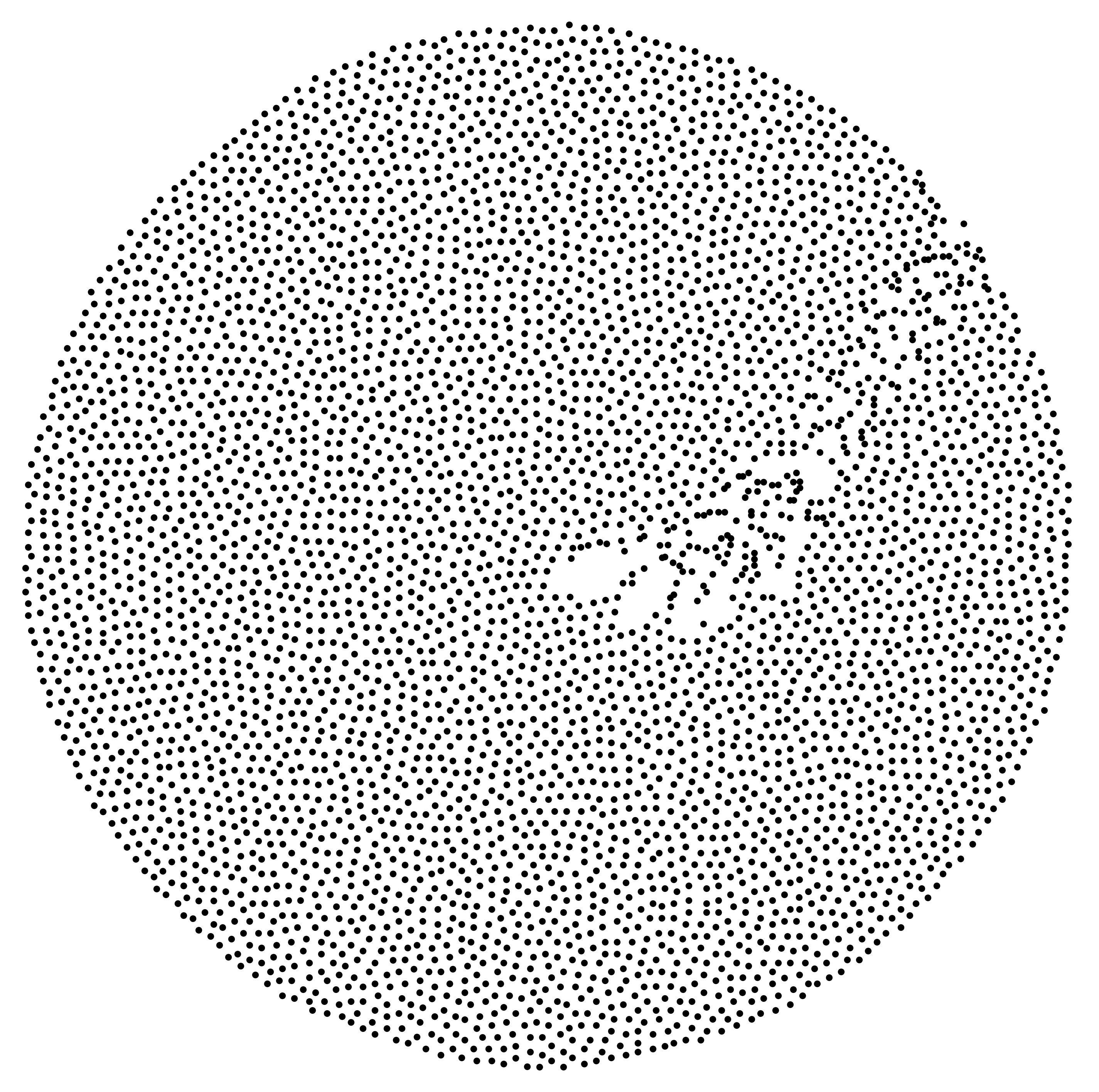}
    \includegraphics[width=0.4\linewidth]{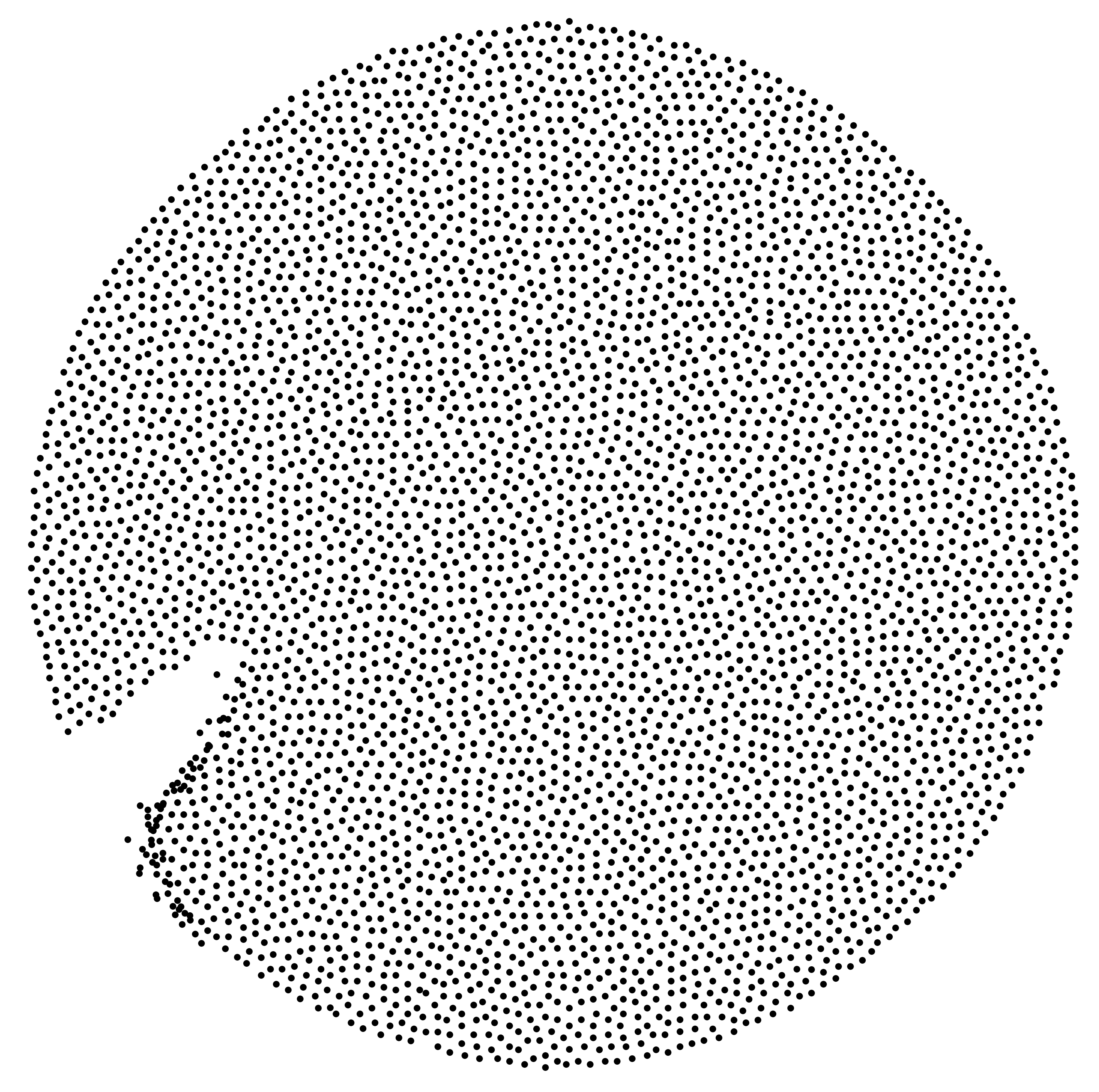}
    \includegraphics[width=0.4\linewidth]{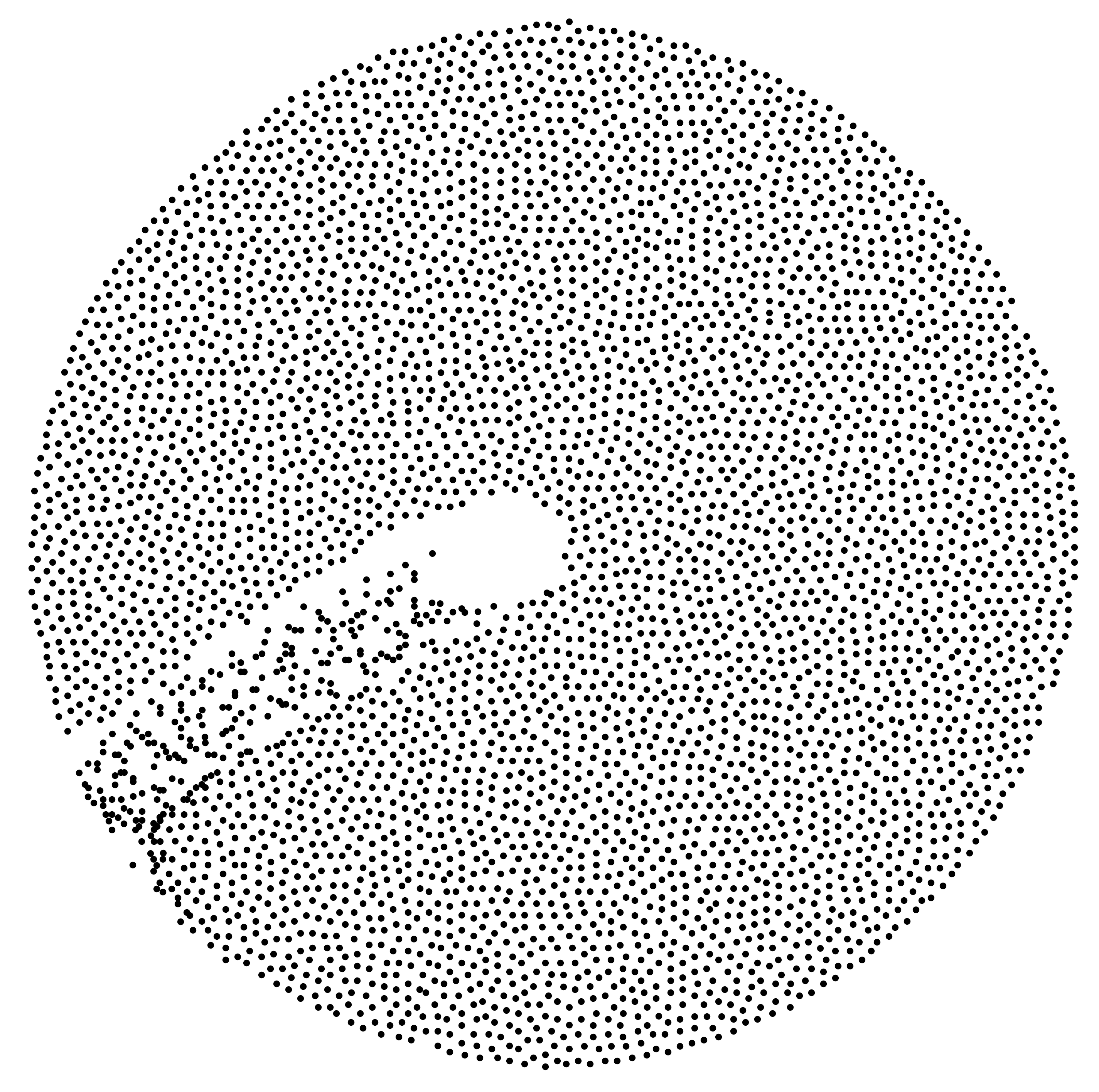}
    \includegraphics[width=0.4\linewidth]{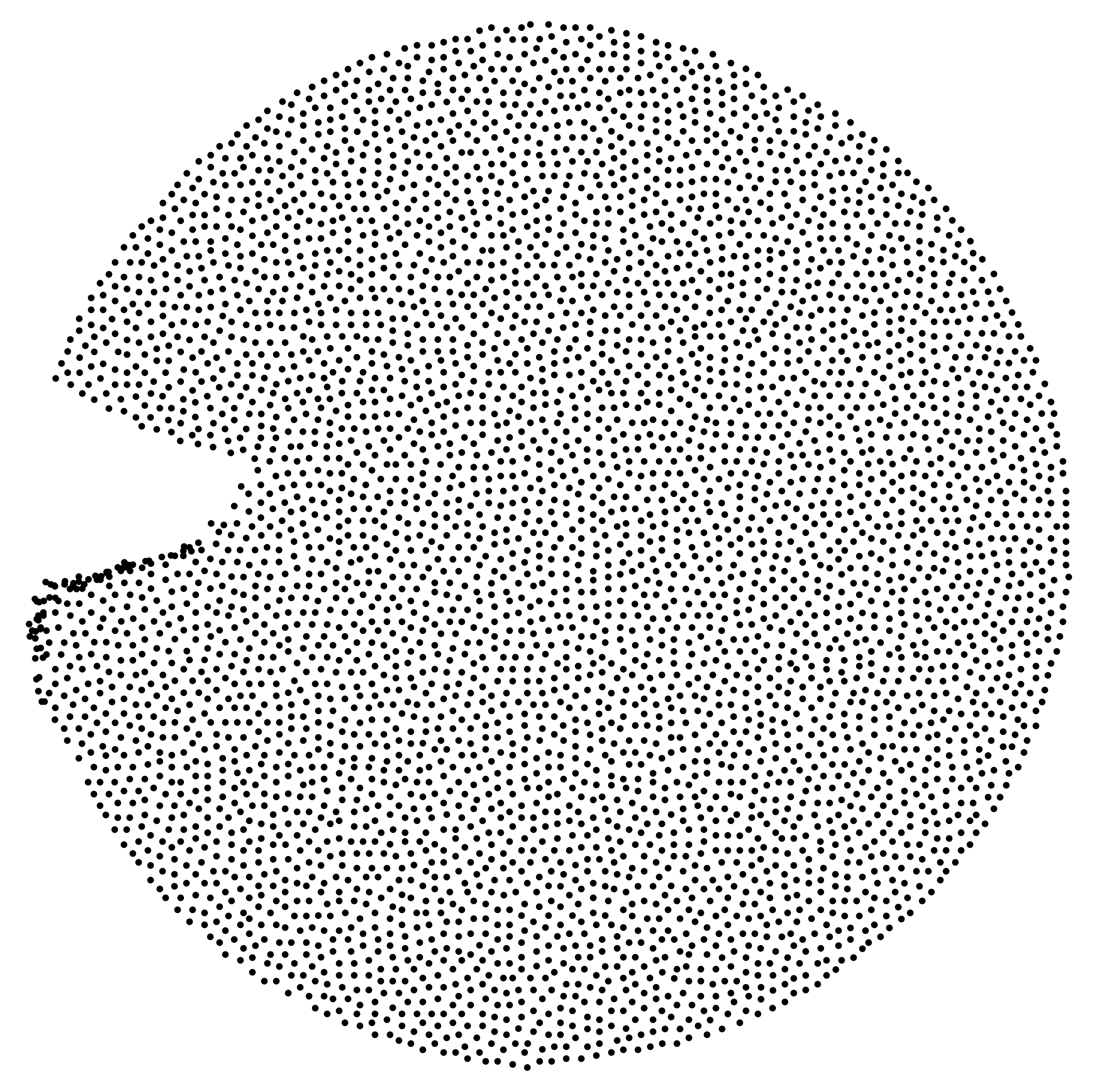}
    \includegraphics[width=0.4\linewidth]{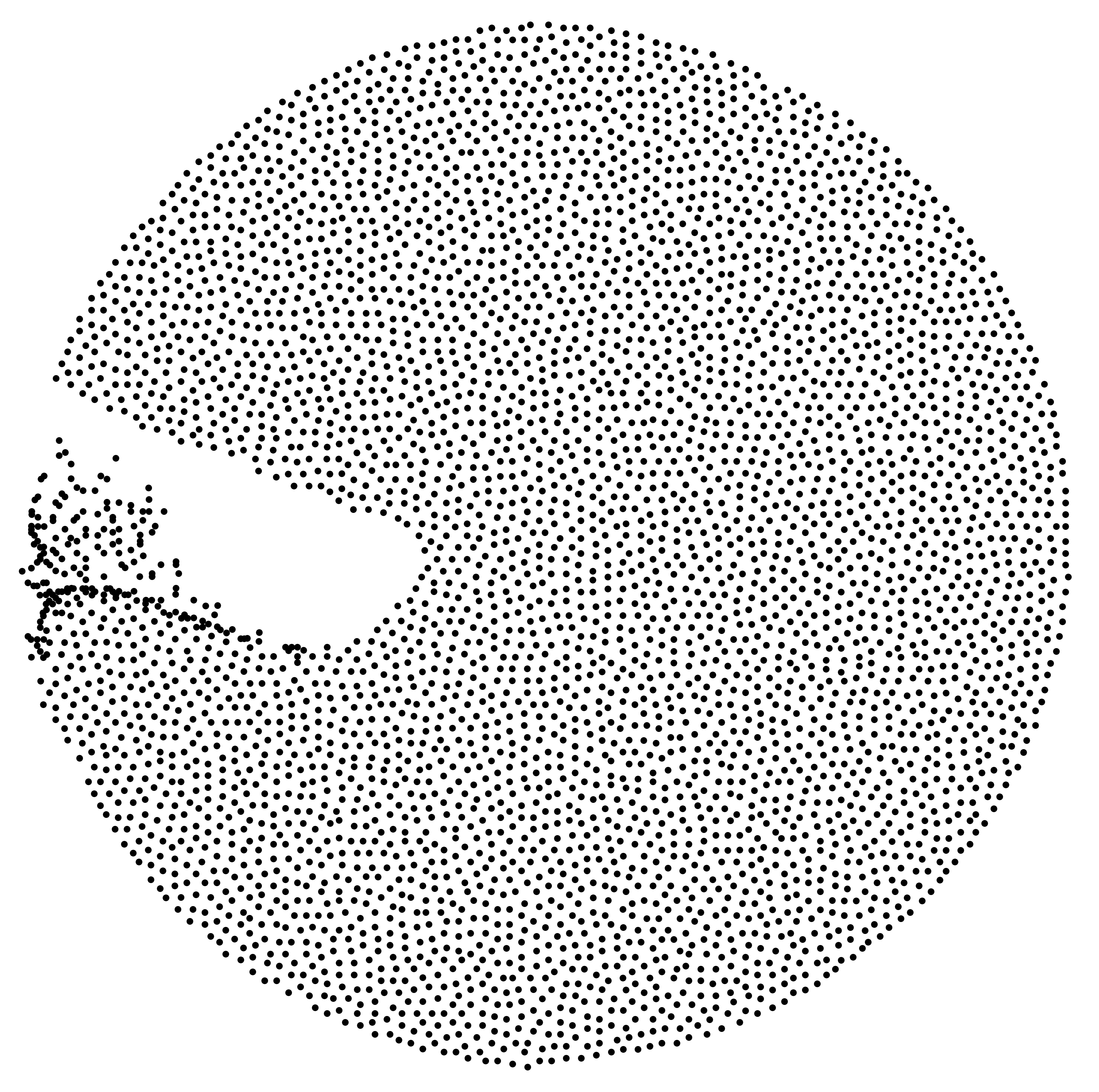}
    \caption{State of the vortex array, as observed in simulations, at the end of the exodus (left) and at the end of the entire event (right), arranged from top to bottom, increasing in $v_{\text{crit}}$ (30, 60, 90). Black dots represent vortices. Pinning sites, more numerous than the vortices, are not shown. In all the cases, the radius of the container R=10. The opening angle $\phi^{*}$ and the depth $d^{*}$ recorded at the end of exodus increases with $v_{\text{crit}}$. However, the total depth of the disturbance, apparent at the end of the entire event, is seen to reduce with $v_{\text{crit}}$.}
    \label{fig:2}
    \vspace{-4pt}
\end{figure}

\textit{Superfluid simulations.--} Using a previously established setup \cite{howitt2020, anantharaman2026}, we simulate the cross section of a spinning-down cylindrical container of radius $R = 10$, moment of inertia $I_{c} = 1$, and having $10^5$ identical pinning sites arranged in a square lattice. The container, initially rotating with an angular speed $\Omega_{0} = 50$, houses a superfluid of moment of inertia $I_s = 1$, having $N = 5\times10^3$ pinned point vortices, each endowed with a quantized circulation of $\kappa = 2 \pi$. Here, we employ a dimensionless coordinate system where the length unit is one, the time unit is constructed from $\kappa$, and the mass unit is derived from $I_{c}$. All other quantities are defined through this fundamental set. The details of the simulator, including the implementation of pinning, are provided in the Supplemental Material (SM). Initially, the superfluid and the container are in synchrony. At time $t=0$, linear spin-down is turned on, after which the state of the vortex array and the rotation rate of the container are both tracked continuously. We run three such simulations, differing only in the critical unpinning speed $v_{\text{crit}}$ characterizing the pinning sites. The chosen values of $v_{\text{crit}}$ (30, 60, 90) ensure that the critical opening angle suggested by Eq.~(2) is much larger than the minimum angular resolution provided by the finite array, given by the ratio of the average inter-vortex spacing to the container's radius. Fig. \ref{fig:2} shows, for the three simulations: a) the state of the array at the end of the first exodus, estimated by studying the velocity field at the boundary, and b) the state at the end of the entire rearrangement. See SM for animations of the rearrangement in each case, juxtaposed with the evolution of the velocity field across the superfluid. The characteristic scales of the depletion region are evident. Also notice that the post-exodus rearrangement widens the tip and alters the depth to which vortices are disturbed. We extract the values of $\phi^{*}$ and $d^{*}$ from the configuration at the end of the exodus, and the value of $\Delta \Omega_{c}^{*}/\Omega_{0}$ as the fractional spin-up recorded at the end of the entire event. In Table \ref{tab:1}, we compare the same with estimates obtained from the analytical formulation \footnotetext{We use Eq.~(\ref{eq:18}) derived in the End Matter, instead of the simplified form presented in Eq.~(\ref{eq:3}), to obtain the relevant values of $d^{*}$, since the opening angles involved in our simulations are not necessarily small.} and find a correspondence between the two. This suggests that the contribution of the post-exodus rearrangement to the net transfer of angular momentum is marginal for sectors with small opening angles, validating the mechanisms proposed in the earlier sections.

\begin{table}[hb]
\caption{\label{tab:1} Comparison of the characteristic scales and the fractional spin-ups obtained from simulations with those estimated by analysis, for three cases varying only in the critical unpinning speed $v_{\text{crit}}$. The subscript $_{sim}$ indicates quantities relevant to simulations. All units are as mentioned in the main text.}
\begin{ruledtabular}
\begin{tabular}{ccccccc}
 $v_{\text{crit}}$ &$\phi_{sim}^{*}$ &$d_{sim}^{*}$ &$\Delta\Omega_{c,sim}^{*}/\Omega_{0}$ &$\phi^{*}$ &$d^{*}$\cite{Note1} &$\Delta\Omega_{c}^{*}/\Omega_{0}$\\
\hline\\
30 &$0.350$ &3.200 &0.015 &0.280 &3.140 &0.013 \\
60 &$0.590$ &4.000 &0.039 &0.570 &4.320 &0.041 \\
90 &$0.870$ &5.100 &0.049 &0.850 &5.000 &0.082 \\
\end{tabular}
\end{ruledtabular}
\end{table}

\textit{Pulsar glitches.--} Pulsars are neutron stars undergoing spin-down primarily due to magnetic dipole radiation. Some pulsars exhibit sudden spin-ups, referred to as glitches, with the rotation relaxing back to the original rate over a span of months to years, suggesting superfluid presence \cite{manchester1990, antonopoulou2022}. The inner crust and the outer core of the star are thought to contain neutron superfluid in the singlet and the triplet state, respectively. The vortices in the former region are expected to pin to nuclear lattice sites, thus decoupling the superfluid rotation from that of the crust. Vortices in the core, on the other hand, are strongly magnetized due to the entrainment of superconducting protons. This allows the superfluid to corotate with the crust via scattering of electrons. However, the superconducting protons in this region also cause the magnetic field to be configured as an array of flux tubes, entities that could serve as pinning centres \cite{sauls1989}. Despite the structural uncertainties, parallels with the previous discussions are clear, that both quantized vortices and pinning structures likely span a large volume of the spinning-down star, presenting the possibility of sudden vortex rearrangements. Unresolved, in the study of pulsar glitches, are the conditions that determine their onset, progress, and termination. We probe the relevance of the mechanism elucidated in the previous sections to pulsar glitches by estimating the fractional spin-up $\Delta\Omega_{c}^{*}/\Omega_{0}$ for two pulsars, the Crab and the Vela, and comparing the same with observed fractional glitch sizes. We approximate the star as a cylindrical container of radius $R$ rotating at a rate $\Omega_{0}$ and having a large number of identical pinning sites characterized by a critical unpinning velocity $v_{\text{crit}}$. Further, let the reservoir of superfluid participating in a glitch have a moment of inertia $I_s$, with the rest of the star having $I_c$. The radius of neutron stars is well-constrained around $\rm{10~km}$, the rotation rate and the fractional moment of inertia $I_s/I_c$ have both been inferred from observations, and the critical unpinning velocities across the crust have been estimated by considering microscopic nuclear interactions at various densities and also by applying vortex creep theory to the observed recovery of glitches \cite{link1999, espinoza2011, basu2022, alpar1984a, alpar1984b, seveso2016}. In Table \ref{tab:2}, we gather the relevant values for both the Crab and the Vela pulsars, alongside the estimates of $\phi^{*}$, $d^{*}$, and $\Delta\Omega_{c}^{*}/\Omega_{0}$. The estimated values suggest fractional spin-ups of $\sim 10^{-10}$ for Crab, and $\sim 10^{-5}$ for Vela. \\

\begin{table}[hb]
\caption{\label{tab:2}Estimates of the characteristic scales and the fractional spin-ups for two representative pulsars, the Crab and the Vela, tabulated along with the relevant stellar parameters. The units are mentioned in the header.}
\begin{ruledtabular}
\begin{tabular}{cccccccc}
 Pulsar &$R$ &$\Omega_{0}$ &$I_s/I_c$ &$v_{\text{crit}}$ &$\phi^{*}$ &$d^{*}$ &$\Delta\Omega_{c}^{*}/\Omega_{0}$\\
  &$\rm{km}$ &$\rm{rad~s^{-1}}$ &$-$ &$\rm{km~s^{-1}}$ &$\rm{10^{-4}~rad}$ &$\rm{km}$ &$-$\\
\hline\\
Crab &10 &188 &$10^{-4}$ &0.1 &$2.5$ &0.7 &$\sim10^{-10}$ \\
Vela &10 &70 &$10^{-2}$ &10 &$670.0 $ &1.9 &$\sim 10^{-5}$ \\
\end{tabular}
\end{ruledtabular}
\end{table}

\noindent While both pulsars have exhibited a range of glitch sizes, most Crab glitches have a fractional change of about $10^{-9}$, whereas most in Vela are observed to be around $10^{-6}$, close to the obtained estimates \cite{basu2022}. The simple picture established here indicates that pulsar glitches could vacate regions extending to depths larger than that of the crust ($\rm{\sim 1~km}$). The relevance of this suggestion to real pulsars, having a spherical structure with radially stratified physical properties, needs deliberation.

\textit{Conclusions and discussion.--} Pinned spinning-down superfluids experience a sudden reconfiguration of their vortex array, reflecting as a spin-up in the container's rate of rotation. We show this to be a self-arresting process, having a well-defined spatial extent, whose dynamics is dictated primarily by the evolution of the velocity field, and reinforced by the transfer of angular momentum. The characteristic spin-ups of the Crab and the Vela pulsars suggested by this process roughly agree with observations. While the recasting of a complex system into a simple analytical framework has helped us realize the arresting mechanism of vortex rearrangements, the ideal conditions considered here necessarily make the study only a limiting case. For instance, the evolution of the array in an inhomogeneous pinning landscape, as in neutron stars, is not obvious. However, it is clear that the velocity field of the array at the end of the reconfiguration is non-axisymmetric and characterized by two scales, $\phi^{*}$ and $d^{*}$. This flow remains almost frozen in the container's frame of reference, till the next event. In neutron stars, such features give rise to a current quadrupole moment, holding the potential to produce continuous gravitational waves, a line of inquiry we save for the future \cite{thorne1980, cheunchitra2024}. While the paradigm discussed in this work includes qualitative features of both instabilities and avalanches, it is not straightforward to categorize it under either umbrella. A linear instability generally refers to the exponential growth of an infinitesimal perturbation \cite{landau1987}. Whereas, an avalanche, illustrated by the BTW sand-pile in its critical state, exhibits no characteristic scales \cite{bak1987}. The phenomenon explored here, however, does not amplify any arbitrary perturbation and involves rearrangements of well-defined angular and radial extent. To underscore this difference, we term this process \textit{allelo-kinesis}, referring to dynamics dictated by mutual interactions.

\textit{Acknowledgements.--} This article benefited from discussions within COST Action SCALES, CA24139, supported by COST (European Cooperation in Science and Technology). The authors thank Ashoka University (AU) and Axis Bank for continued support. The simulations were run on two different machines at AU; Aryabhatta (courtesy Prof. Suratna Das, access facilitated by Umang Kumar), and Tao (courtesy Prof. Garima Mishra, facilitated by Souradeep Sengupta and Bidisha Mukherjee). The supplemental animations were prepared using Matplotlib, a Python package \cite{matplotlib2007}.

\textit{Data availability.--} The simulation data used in this article can be made available upon request by emailing the corresponding author. However, the codes developed by the authors have been made publicly available through the following repository: \href{https://github.com/ananth-94/vortexsimulator}{github.com/ananth-94/vortexsimulator}. They can be used by citing the article where they were introduced \cite{anantharaman2026}.

\bibliography{references}

@misc{perkins2026,
  author = {Perkins, C. and Igoshev, A. and Wood, T.},
  title  = {{Hydrodynamic Simulation of Glitches in Neutron Star Crusts}},
  note   = {Presented at SCALES 2026, Coimbra, Portugal},
  month  = {June},
  year   = {2026},
  url = {https://indico.global/event/17311/contributions/156506/}
}

@article{alpar1984a,
  title = {Vortex Creep and the Internal Temperature of Neutron Stars. {{I}} - {{General}} Theory},
  author = {Alpar, M. A. and Pines, D. and Anderson, P. W. and Shaham, J.},
  year = 1984,
  month = jan,
  journal = {The Astrophysical Journal},
  volume = {276},
  pages = {325},
  issn = {0004-637X, 1538-4357},
  doi = {10.1086/161616},
  urldate = {2026-11-13},
  langid = {english}
}

@article{alpar1984b,
  title = {Vortex Creep and the Internal Temperature of Neutron Stars. {{II}} - {{VELA}} Pulsar},
  author = {Alpar, M. A. and Anderson, P. W. and Pines, D. and Shaham, J.},
  year = 1984,
  month = mar,
  journal = {The Astrophysical Journal},
  volume = {278},
  pages = {791},
  issn = {0004-637X, 1538-4357},
  doi = {10.1086/161849},
  urldate = {2026-11-13},
  langid = {english}
}

@article{anantharaman2026,
  title = {Pulsar Glitches Due to Triggered Vortex Traps},
  author = {Anantharaman, S V and Bhattacharya, Dipankar and Alpar, M Ali and G{\"u}gercino{\u g}lu, Erbil},
  year = 2026,
  month = nov,
  journal = {Monthly Notices of the Royal Astronomical Society},
  volume = {553},
  number = {2},
  pages = {stag2072},
  issn = {0035-8711, 1365-2966},
  doi = {10.1093/mnras/stag2072},
  urldate = {2026-11-13},
  copyright = {https://creativecommons.org/licenses/by/4.0/},
  langid = {english}
}

@article{anderson1975,
  title = {Pulsar Glitches and Restlessness as a Hard Superfluidity Phenomenon},
  author = {Anderson, P. W. and Itoh, N.},
  year = 1975,
  month = jul,
  journal = {Nature},
  volume = {256},
  number = {5512},
  pages = {25--27},
  issn = {0028-0836, 1476-4687},
  doi = {10.1038/256025a0},
  urldate = {2024-04-11},
  copyright = {http://www.springer.com/tdm},
  langid = {english}
}

@article{antonopoulou2022,
  title = {Pulsar Glitches: Observations and Physical Interpretation},
  shorttitle = {Pulsar Glitches},
  author = {Antonopoulou, Danai and Haskell, Brynmor and Espinoza, Crist{\'o}bal M.},
  year = 2022,
  month = dec,
  journal = {Reports on Progress in Physics},
  volume = {85},
  number = {12},
  pages = {126901},
  publisher = {IOP Publishing},
  issn = {0034-4885},
  doi = {10.1088/1361-6633/ac9ced},
  urldate = {2024-04-07},
  langid = {english}
}

@article{bak1987,
  title = {Self-Organized Criticality: {{An}} Explanation of the 1/f Noise},
  shorttitle = {Self-Organized Criticality},
  author = {Bak, Per and Tang, Chao and Wiesenfeld, Kurt},
  year = 1987,
  month = jul,
  journal = {Physical Review Letters},
  volume = {59},
  number = {4},
  pages = {381--384},
  publisher = {American Physical Society},
  doi = {10.1103/PhysRevLett.59.381},
  urldate = {2024-09-23}
}

@article{barnes1986,
  title = {A Hierarchical {{O}}({{N}} Log {{N}}) Force-Calculation Algorithm},
  author = {Barnes, Josh and Hut, Piet},
  year = 1986,
  month = dec,
  journal = {Nature},
  volume = {324},
  number = {6096},
  pages = {446--449},
  publisher = {Nature Publishing Group},
  issn = {1476-4687},
  doi = {10.1038/324446a0},
  urldate = {2025-05-03},
  copyright = {1986 Springer Nature Limited},
  langid = {english}
}

@article{basu2022,
  title = {The {{Jodrell}} Bank Glitch Catalogue: 106 New Rotational Glitches in 70 Pulsars},
  shorttitle = {The {{Jodrell}} Bank Glitch Catalogue},
  author = {Basu, A and Shaw, B and Antonopoulou, D and Keith, M J and Lyne, A G and Mickaliger, M B and Stappers, B W and Weltevrede, P and Jordan, C A},
  year = 2022,
  month = mar,
  journal = {Monthly Notices of the Royal Astronomical Society},
  volume = {510},
  number = {3},
  pages = {4049--4062},
  issn = {0035-8711},
  doi = {10.1093/mnras/stab3336},
  urldate = {2024-09-22}
}

@article{cheunchitra2024,
  title = {Persistent Gravitational Radiation from Glitching Pulsars -- {{II}}. {{Updated}} Scaling with Vortex Number},
  author = {Cheunchitra, T and Melatos, A and Carlin, J B and Howitt, G},
  year = 2024,
  month = jan,
  journal = {Monthly Notices of the Royal Astronomical Society},
  volume = {528},
  number = {2},
  pages = {1360--1371},
  issn = {0035-8711, 1365-2966},
  doi = {10.1093/mnras/stae130},
  urldate = {2024-05-15},
  copyright = {https://creativecommons.org/licenses/by/4.0/},
  langid = {english}
}

@book{donnelly1991,
  title = {Quantized Vortices in Helium {{II}}},
  author = {Donnelly, Russell J.},
  year = 1991,
  series = {Cambridge Studies in Low Temperature Physics},
  number = {3},
  publisher = {Cambridge University Press},
  isbn = {978-0-521-32400-7},
  lccn = {QC175.4 .D58 1991},
  url = {https://www.cambridge.org/in/universitypress/subjects/mathematics/fluid-dynamics-and-solid-mechanics/quantized-vortices-helium-ii?format=HB&isbn=9780521324007}
}

@article{espinoza2011,
  title = {A Study of 315 Glitches in the Rotation of 102 Pulsars},
  author = {Espinoza, C. M. and Lyne, A. G. and Stappers, B. W. and Kramer, M.},
  year = 2011,
  month = jun,
  journal = {Monthly Notices of the Royal Astronomical Society},
  volume = {414},
  number = {2},
  pages = {1679--1704},
  issn = {0035-8711},
  doi = {10.1111/j.1365-2966.2011.18503.x},
  urldate = {2024-04-09}
}

@article{fetter1965,
  title = {Vortices in an {{Imperfect Bose Gas}}. {{I}}. {{The Condensate}}},
  author = {Fetter, Alexander L.},
  year = 1965,
  month = apr,
  journal = {Physical Review},
  volume = {138},
  number = {2A},
  pages = {A429-A437},
  issn = {0031-899X},
  doi = {10.1103/PhysRev.138.A429},
  urldate = {2026-10-05},
  copyright = {http://link.aps.org/licenses/aps-default-license},
  langid = {english}
}

@incollection{feynman1955,
  title = {Chapter {{II Application}} of {{Quantum Mechanics}} to {{Liquid Helium}}},
  booktitle = {Progress in {{Low Temperature Physics}}},
  author = {{R. P. Feynman}},
  year = 1955,
  volume = {1},
  pages = {17--53},
  publisher = {Elsevier},
  doi = {10.1016/S0079-6417(08)60077-3},
  urldate = {2026-11-13},
  copyright = {https://www.elsevier.com/tdm/userlicense/1.0/},
  isbn = {978-0-444-53307-4},
  langid = {english}
}

@article{graber2017,
  author  = {Graber, Vanessa and Andersson, Nils and Hogg, Michael},
  title   = {Neutron stars in the laboratory},
  journal = {Int. J. Mod. Phys. D},
  year    = {2017},
  volume  = {26},
  pages   = {1730015},
  doi     = {10.1142/S0218271817300154}
}

@article{howitt2020,
  title = {Simulating Pulsar Glitches: An {{N}}-Body Solver for Superfluid Vortex Motion in Two Dimensions},
  shorttitle = {Simulating Pulsar Glitches},
  author = {Howitt, G. and Melatos, A. and Haskell, B.},
  year = 2020,
  month = oct,
  journal = {Monthly Notices of the Royal Astronomical Society},
  volume = {498},
  number = {1},
  primaryclass = {astro-ph, physics:cond-mat},
  pages = {320--331},
  issn = {0035-8711, 1365-2966},
  doi = {10.1093/mnras/staa2314},
  urldate = {2024-04-07}
}

@book{landau1987,
  title = {{Fluid mechanics}},
  author = {L. D. Landau and E. M. Lifshitz},
  year = 1987,
  series = {{Course of theoretical physics}},
  edition = {2nd ed., 2nd English ed., rev},
  number = {v. 6},
  publisher = {Pergamon Press},
  address = {Oxford, England ; New York},
  isbn = {978-0-08-033933-7 978-0-08-033932-0},
  langid = {eng rus},
  lccn = {QA901 .L283 1987}
}

@article{link1999,
  title = {Pulsar {{Constraints}} on {{Neutron Star Structure}} and {{Equation}} of {{State}}},
  author = {Link, Bennett and Epstein, Richard I. and Lattimer, James M.},
  year = 1999,
  month = oct,
  journal = {Physical Review Letters},
  volume = {83},
  number = {17},
  pages = {3362--3365},
  issn = {0031-9007, 1079-7114},
  doi = {10.1103/PhysRevLett.83.3362},
  urldate = {2026-11-13},
  copyright = {http://link.aps.org/licenses/aps-default-license},
  langid = {english}
}

@article{liu2024,
  title = {Vortex {{Avalanches}} and {{Collective Motion}} in {{Neutron Stars}}},
  author = {Liu, I-Kang and Baggaley, Andrew W. and Barenghi, Carlo F. and Wood, Toby S.},
  year = 2025,
  month = apr,
  journal = {The Astrophysical Journal},
  volume = {984},
  number = {1},
  pages = {83},
  publisher = {The American Astronomical Society},
  issn = {0004-637X},
  doi = {10.3847/1538-4357/adc383},
  urldate = {2025-05-06},
  langid = {english}
}

@article{manchester1990,
  title = {Pulsars Miss a Beat},
  author = {Manchester, R. N.},
  year = 1990,
  month = jan,
  journal = {Nature},
  volume = {343},
  number = {6256},
  pages = {317--318},
  publisher = {Nature Publishing Group},
  issn = {1476-4687},
  doi = {10.1038/343317a0},
  urldate = {2024-09-02},
  copyright = {1990 Springer Nature Limited},
  langid = {english}
}

@article{matplotlib2007,
  Author    = {Hunter, J. D.},
  Title     = {Matplotlib: A 2{{D}} graphics environment},
  Journal   = {Computing in Science \& Engineering},
  Volume    = {9},
  Number    = {3},
  Pages     = {90--95},
  doi       = {10.1109/MCSE.2007.55},
  year      = {2007}
}

@article{onsager1949,
  title = {Statistical Hydrodynamics},
  author = {Onsager, L.},
  year = 1949,
  month = mar,
  journal = {Il Nuovo Cimento},
  volume = {6},
  number = {S2},
  pages = {279--287},
  issn = {0029-6341, 1827-6121},
  doi = {10.1007/BF02780991},
  urldate = {2026-11-13},
  copyright = {http://www.springer.com/tdm},
  langid = {english}
}

@article{poli2023,
  title = {Glitches in {{Rotating Supersolids}}},
  author = {Poli, Elena and Bland, Thomas and White, Samuel J. M. and Mark, Manfred J. and Ferlaino, Francesca and Trabucco, Silvia and Mannarelli, Massimo},
  year = 2023,
  month = nov,
  journal = {Physical Review Letters},
  volume = {131},
  number = {22},
  pages = {223401},
  issn = {0031-9007, 1079-7114},
  doi = {10.1103/PhysRevLett.131.223401},
  urldate = {2025-01-02},
  langid = {english}
}

@incollection{sauls1989,
  title = {Superfluidity in the {{Interiors}} of {{Neutron Stars}}},
  booktitle = {Timing {{Neutron Stars}}},
  author = {Sauls, J. A.},
  editor = {{\"O}gelman, H. and {van den Heuvel}, E. P. J.},
  year = 1989,
  pages = {457--490},
  publisher = {Springer Netherlands},
  address = {Dordrecht},
  doi = {10.1007/978-94-009-2273-0_43},
  urldate = {2024-09-25},
  isbn = {978-94-009-2273-0},
  langid = {english}
}

@article{seveso2016,
  title = {Mesoscopic Pinning Forces in Neutron Star Crusts},
  author = {Seveso, S. and Pizzochero, P. M. and Grill, F. and Haskell, B.},
  year = 2016,
  month = feb,
  journal = {Monthly Notices of the Royal Astronomical Society},
  volume = {455},
  number = {4},
  pages = {3952--3967},
  issn = {0035-8711, 1365-2966},
  doi = {10.1093/mnras/stv2579},
  urldate = {2026-11-13},
  langid = {english}
}

@article{skrbek2026,
  title = {Spin-down of a Quantum Fluid and Neutron Stars' Glitches: {{The}} Experiments of {{J}}. {{S}}. {{Tsakadze}} and {{S}}. {{E}}. {{Tsakadze}} Revisited},
  shorttitle = {Spin-down of a Quantum Fluid and Neutron Stars' Glitches},
  author = {Skrbek, L. and Barenghi, Carlo F.},
  year = 2026,
  month = nov,
  journal = {Physical Review Fluids},
  volume = {11},
  number = {11},
  pages = {114701},
  publisher = {American Physical Society},
  doi = {10.1103/2pl2-xz63},
  urldate = {2026-11-13}
}

@article{thorne1980,
  title = {Multipole Expansions of Gravitational Radiation},
  author = {Thorne, Kip S.},
  year = 1980,
  month = apr,
  journal = {Reviews of Modern Physics},
  volume = {52},
  number = {2},
  pages = {299--339},
  publisher = {American Physical Society},
  doi = {10.1103/RevModPhys.52.299},
  urldate = {2026-11-13}
}

@article{tsakadze1980,
  title = {Properties of Slowly Rotating Helium {{II}} and the Superfluidity of Pulsars},
  author = {Tsakadze, J. S. and Tsakadze, S. J.},
  year = 1980,
  month = jun,
  journal = {Journal of Low Temperature Physics},
  volume = {39},
  number = {5},
  pages = {649--688},
  issn = {1573-7357},
  doi = {10.1007/BF00114899},
  urldate = {2026-10-10},
  langid = {english}
}

@article{warszawski2011,
  title = {Gross--{{Pitaevskii}} Model of Pulsar Glitches},
  author = {Warszawski, L. and Melatos, A.},
  year = 2011,
  month = aug,
  journal = {Monthly Notices of the Royal Astronomical Society},
  volume = {415},
  number = {2},
  pages = {1611--1630},
  issn = {0035-8711},
  doi = {10.1111/j.1365-2966.2011.18803.x},
  urldate = {2024-08-13}
}

\appendix

\section{End Matter}

\subsection{Velocity field at the tip for small sector depths}

The first unpinning leads to a small sector, of depth $d$ and opening angle $\phi$, devoid of vortices, while the rest of the array remains pinned. The superfluid velocity field has contributions from the lag generated due to spin-down, the removed sector, and the corresponding image sector. At the tip, the velocity lag due to spin-down is
\begin{equation}
    \mathbf{v_{tip,~lag}} = v_{\text{crit}}\frac{R-d}{R}~\mathbf{\hat{\theta}}.
    \label{eq:9}
\end{equation}
The contribution of the removed sector is found by integrating the effect of its individual array elements, having counter-rotating vortices of areal density $n_{v}$. Here, we consider both arms of the sector to have the same length $d$, and the opening angle to be small. Away from the X-axis, this geometry becomes incommensurate with the actual boundary of the container. However, this only yields a higher order correction to the field at the tip, an effect we do not consider below. Using the Onsager-Feynman relation to express the net circulation per unit area ($n_{v}\kappa$) in terms of the initial rotation rate, we obtain
\begin{equation}
\begin{aligned}
    \mathbf{v_{tip,~sec}} &= - n_{v}\int_{0}^{\phi}d\phi^{\prime} \int_{0}^{d}\left[\left(\frac{-\kappa}{2 \pi r^{\prime}}\right) \cos{\phi^{\prime}}\right] r^{\prime} dr^{\prime}~\mathbf{\hat{\theta}}\\
    &= \frac{\Omega_{0}d}{\pi}\phi~\mathbf{\hat{\theta}},
\end{aligned}
\label{eq:10}
\end{equation}
where $r^{\prime}$ is the distance from the tip to an arbitrary element and $\phi^{\prime}$ is the angle subtended with respect to the positive X-axis. Additionally, the container wall allows no radial flow of superfluid. This condition is effected by introducing image vortices outside the boundary, such that an array element at radius $r$ and angle $\theta$, with respect to the centre of the container, has its corresponding image, with opposite circulation, placed at $R^{2}/r$ and $\theta$. In the presence of a depleted region, to maintain the correct boundary condition, the corresponding image vortices also need to be removed. The resultant effect at the tip can be found by exact integration. However, the leading order contribution of the image sector may be obtained by positioning the vortices at its centroid. For small sector depths and opening angle, note that the image is almost a trivial reflection about the boundary. In such a case, we may approximate all the vortices to be on the X-axis, collinear with the tip, at a distance $d/2$ outside the container wall, giving us
\begin{equation}
\begin{aligned}
    \mathbf{v_{tip,~img}} &= - n_{v} \frac{\kappa}{2 \pi (3d/2)}\int_{0}^{\phi} d\phi^{\prime} \int_{0}^{d} r^{\prime} dr^{\prime}~\mathbf{\hat{\theta}}\\
    &= -\frac{\Omega_{0}d}{3\pi}\phi~\mathbf{\hat{\theta}}.
\end{aligned}
\label{eq:11}
\end{equation}
The net tangential flow at the tip is then (cf. Eq.~(\ref{eq:1}))
\begin{equation*}
    \mathbf{v_{tip}} = \left[v_{\text{crit}}\frac{R-d}{R}+\frac{\Omega_{0}d}{\pi}\phi-\frac{\Omega_{0}d}{3\pi} \phi \right] \mathbf{\hat{\theta}}.
\end{equation*}

\subsection{Velocity field at the wall for moderate sector depths}

The flow at the wall incurs three contributions, similar to the tip, with the velocity lag due to spin-down having the simple form
\begin{equation}
    \mathbf{v_{wall,~lag}} = v_{\text{crit}}~\mathbf{\hat{\theta}}.
    \label{eq:12}
\end{equation} 
Approximating the removed sector as a right triangle, the hypotenuse aligning with the forward arm, allows us to integrate its effect as
\begin{equation}
\begin{aligned}
    \mathbf{v_{wall,~sec}} = ~&n_{v}\int_{0}^{\alpha}d\alpha^{\prime} \int_{0}^{y~/\cos{\alpha^{\prime}}}\Biggl[\left(\frac{-\kappa}{2 \pi r^{\prime}}\right).\\
    &\sin\left({\alpha^{\prime}}+\tan^{-1}\left(\frac{d}{R}\tan{\phi}\right)\right)\Biggr] r^{\prime} dr^{\prime}~\mathbf{\hat{\theta}},
\end{aligned}
\label{eq:13}
\end{equation}
where $r^{\prime}$ is the distance from the wall to an arbitrary element within the vacated region, $\alpha^{\prime}$ is the angle subtended by the element with respect to the vertical of the triangle (perpendicular to the X-axis), $y \sim d(\tan \phi)$ is the height of the vertical, and $\alpha = \frac{\pi}{2} - \phi$ is the angle subtended by the forward arm with respect to the vertical. The total flow induced at the wall by any given element has a component along the tangent, the projection being captured by the sinusoidal term. Evaluating the integral lends
\begin{equation}
    \mathbf{{v}_{wall,~sec}} =- \frac{\Omega_{0} d}{\pi}\tan\phi \Biggl(\frac{d}{R}\phi\left(\frac{\pi}{2}-\phi\right)-\ln{\left(\sin\phi\right)}\Biggr) ~\mathbf{\hat{\theta}},
\label{eq:14}
\end{equation}
where the term quadratic in $d$ arises due to the curvature of the boundary. For sectors with moderate and large depths, the corresponding image can no longer be treated as a simple reflection. However, the leading order contribution may still be obtained by situating all the vortices at the centroid. The tip of the real sector has its image on the X-axis, at a distance $Rd/(R-d)$ from the boundary, forming the base of the image triangle. The height of the vertical is $\sim d\tan{\phi}$. The field due to the image patch is then
\begin{equation}
\begin{aligned}
    \mathbf{v_{wall,~img}} = - &\left(n_{v}\frac{d^2}{2}\tan{\phi}\right) \frac{\kappa}{2 \pi r_{cent}}.\\
    &\sin\left({\beta_{cent}}\right) \cos\left(\tan^{-1}\left(\frac{d}{R}\tan{\phi}\right)\right)~\mathbf{\hat{\theta}},
\end{aligned}
\label{eq:15}
\end{equation}
where $r_{cent}$ is the distance from the wall to the centroid, and $\beta_{cent}$ is the angle subtended with respect to the vertical. The first bracketed term is the total number of vortices in the sector, whereas the product of the sine and the cosine terms project the total flow induced at the wall along the local tangent. The angle subtended by the centroid
\begin{equation}
\begin{aligned}
    \beta_{cent} = \tan^{-1}\left(\frac{1}{2\left(1-d/R\right)\tan{\phi}}\right)
\end{aligned}
\label{eq:16}
\end{equation}
simplifies for moderate depths, where $2\left(1-d/R\right) \sim 1$, allowing us to recast Eq.~(\ref{eq:15}) as
\begin{equation}
\begin{aligned}
    \mathbf{v_{wall,~img}} = - \frac{\Omega_{0} d}{\pi} \left(\frac{3}{8}\sin\left(2 \tan^{-1}\frac{1}{\tan\phi}\right)\right)~\mathbf{\hat{\theta}}.
\end{aligned}
\label{eq:17}
\end{equation}
The net tangential flow at the wall is, then
\begin{equation}
\begin{aligned}
    \mathbf{{v}_{wall}} = \Biggl[&v_{\text{crit}}- \frac{\Omega_{0} d}{\pi}\tan\phi \Biggl(\frac{d}{R}\phi\left(\frac{\pi}{2}-\phi\right)-\ln{\left(\sin\phi\right)}\Biggr)-\\
    &\frac{\Omega_{0} d}{\pi}\frac{3}{8}\sin\left(2 \tan^{-1}\frac{1}{\tan\phi}\right)\Biggr]~\mathbf{\hat{\theta}}.
\end{aligned}
\label{eq:18}
\end{equation}
For $\phi \ll 1$, this reduces to (cf. Eq.(\ref{eq:3}))
\begin{equation*}
\begin{aligned}
    \mathbf{{v}_{wall}} = \Biggl[&v_{\text{crit}}- \frac{\Omega_{0} d}{\pi}\phi \Biggl(\frac{d}{R}\phi\left(\frac{\pi}{2}-\phi\right)-\ln{\left(\phi\right)}\Biggr)-\\
    &\frac{\Omega_{0} d}{\pi}\phi \left(\frac{3(1-\phi^{2})}{4}\right)\Biggr]~\mathbf{\hat{\theta}}.
\end{aligned}
\end{equation*}

\clearpage
\setcounter{equation}{0}
\setcounter{figure}{0}
\setcounter{table}{0}
\renewcommand{\theequation}{S\arabic{equation}}
\renewcommand{\thefigure}{S\arabic{figure}}
\renewcommand{\thetable}{S\Roman{table}}

\section{Supplemental Material}

\setcounter{page}{1}
\renewcommand{\thepage}{SM\arabic{page}}

\subsection{Superfluid simulator}

A numerical prescription to study the evolution of pinned superfluids, in a cylindrical container undergoing spin-down, was introduced in \cite{howitt2020}. We use an enhanced version of the same that utilizes the Barnes-Hut scheme to cluster groups of vortices into an effective vortex placed at their centre of mass, significantly boosting the computational performance \cite{anantharaman2026, barnes1986}. This two-dimensional formalism hinges on treating vortices as point-like entities dictating the velocity field of the superfluid, while simultaneously responding to the local flow. Numerous pinning centres, corotating with the container and having localized influence on the vortices, are arranged in a uniform square lattice. Additionally, a minor dissipation is endowed to vortices that move relative to the container, while the boundary is imposed through image vortices. For a superfluid having N vortices, the velocity field experienced by an arbitrary vortex, in the container's frame, is
\begin{equation}
    \frac{\mathrm{d}}{\mathrm{d} t}\left(\begin{array}{l}
x_{i} \\
y_{i}
\end{array}\right)=\mathcal{R}_{\phi_{\rm{diss}}}\left(\begin{array}{c}
v_{i, x} \\
v_{i, y}
\end{array}\right),
\label{eq:19}
\end{equation}
where
\begin{equation}
\begin{aligned}
    v_{i, x}=&-\sum_{j{ \neq i}} \frac{\kappa y_{i j}}{2 \pi r_{i j}^2}+\sum_{j_{\text{img}}=1}^N \frac{\kappa y_{i j_{\text {img}}}}{2 \pi r_{i j_{\text {img}}}^2}+\Omega_c y_i+\\
    &\frac{v_{\text{crit}}}{\xi e^{-1/2}} \exp \left[-\frac{r_{i,\text{pin}}^2}{2 \xi^2}\right]\left(y_{i}-y_{\text {pin }}\right),
    \label{eq:20}
\end{aligned}
\end{equation}
and
\begin{equation}
\begin{aligned}
    v_{i, y}=&\sum_{j \neq i} \frac{\kappa x_{i j}}{2 \pi r_{i j}^2}-\sum_{j_{\text{img}}=1}^N \frac{\kappa x_{i j_{\text {img}}}}{2 \pi r_{i j_{\text {img}}}^2}-\Omega_c x_i-\\
    &\frac{v_{\text{crit}}}{\xi e^{-1/2}} \exp \left[-\frac{r_{i,\text{pin}}^2}{2 \xi^2}\right]\left(x_{i}-x_{\text {pin }}\right) .
    \label{eq:21}
\end{aligned}
\end{equation}
Here, $x_{i}$ and $y_{i}$ indicate the position of the vortex on the Cartesian plane, calculated with respect to the origin at the centre of the container. $r_{ij}$ is the distance between a given pair of vortices, with $x_{ij}$ and $y_{ij}$ being the respective Cartesian components. $r_{ij_{\text {img}}}$, $x_{ij_{\text {img}}}$, and $y_{ij_{\text {img}}}$ denote similar quantities relevant for a vortex-image pair. $r_{i,\text{pin}}$ is the distance between the vortex under consideration and the nearest pinning site, located at ($x_{\text{pin}}$, $y_{\text{pin}}$). $\kappa$ is the quantum of circulation attached to each vortex, and $\Omega_{c}$ is the angular speed of the container. The terms on the right hand side in Eq.~(\ref{eq:20}) and Eq.~(\ref{eq:21}) account for the effect of other vortices, the presence of a boundary imposed via image vortices, the rotation of the container, and the influence of the pinning sites, in that order, on the vortex under consideration. Each pinning site is described by a Gaussian potential that rotates nearby vortices in a clockwise fashion, opposite to the counter-clockwise effect of one vortex on another. Every site provides a maximum speed $v_{\text{crit}}$ at the characteristic pinning radius $\xi$, beyond which the effect of the site decays rapidly. The rotation matrix, $\mathcal{R}_{\phi_{\rm{diss}}}$, accounts for the dissipative interactions in the system, resulting in a slight radially outward motion of free vortices, described by the small angle $\phi_{\rm{diss}}$ with respect to their fast azimuthal motion. The positions of the vortices in the container are evolved by numerically integrating the above equations, given the initial configuration of the array. We run our simulations in two phases, the relaxation and the spin-down.

\subsection{Relaxation}

Initially, the container is set to rotate at a constant rate $\Omega_{c} = \Omega_{0}$, and $N$ vortices of circulation $\kappa$ are distributed uniformly over a cross-section of radius $R$. The initial number of vortices are chosen such that the superfluid rotation rate suggested by the Onsager-Feynman relation matches the container rotation. This configuration is time evolved according to Eq.~(\ref{eq:19}), maintaining the container rotation rate at $\Omega_{0}$, till all the vortices within the system are initially pinned.

\subsection{Spin-down}

Once the system has relaxed into a pinned configuration, we begin the linear spin-down of the container due to a constant torque $N_{\rm{ext}}$. This moment is marked as t=0 in our simulation. Conserving net angular momentum, the rotation rate of the container evolves according to
\begin{equation}
\frac{d \Omega_c}{d t}=\frac{N_{\mathrm{ext}}}{I_{c}}-\frac{1}{I_c} \frac{d L_s}{dt},
\label{eq:22}
\end{equation}
where $N_{\mathrm{ext}}$ is a negative quantity, and $I_{c}$ is the moment of inertia of the container. The total angular momentum of the superfluid corresponding to a given configuration of vortices is captured by \cite{fetter1965}
\begin{equation}
    L_s = I_s \Omega_{0} \sum_{i=1}^{N}\frac{(R^{2} - r_{i}^{2})}{NR^2/2},
\label{eq:23}
\end{equation}
such that $L_s$ is exactly $I_{s} \Omega_{0}$ for the initially uniform array, where $I_{s}$ is the moment of inertia of the superfluid. Thus, the change in the container's rotation rate, during a small interval of time $\Delta t$, is intimately tied to the change in vortex positions, given by
\begin{equation}
\Delta \Omega_c=\frac{N_{\mathrm{ext}}}{I_{c}}\Delta t-\Omega_{0}\frac{I_s}{I_c} \Delta\left(\sum_{i=1}^{N}\frac{(R^{2} - r_{i}^{2})}{NR^2/2}\right).
\label{eq:24}
\end{equation}
This change in the rotation rate is updated at the end of every time step. Notice that Eq.~(\ref{eq:24}) suggests a linear spin-down of the container when the vortex array is fully pinned, and a spin-up if vortices migrate in the radially outward direction.

\vspace{-12pt}
\subsection{Set-up}

We run three simulations, changing only the maximum pinning velocity $v_{\text{crit}}$. The values of the parameters defined above, and as used in our simulations, are tabulated in Table \ref{tab:S1}. Additionally, we take two measures to respect the conditions of the analysis presented in the main text: 1) The influence of each site ($\xi$) is set to half the average inter-site distance ($a$) mimicking area-filling pinning, and 2) The pinning lattice is activated only after the initial vortex configuration is relaxed to a near-Abrikosov state, ensuring that the pinned array is as close to being uniform as possible before spin-down.

\begin{table}[t]
\caption{\label{tab:S1}Summary of the parameters relevant to the simulations described in this article. The units are as mentioned in the main text.}
\begin{ruledtabular}
\begin{tabular}{l}
\textbf{Standard parameters} \\
\hline\\
Radius of the container, $R = 10$ \\
Number of vortices, $N_v = 5\times10^{3}$ \\
Dissipation angle, $\phi_{\rm{diss}} = 0.1$ \\
Moment of inertia of superfluid, $I_{s} = 1$ \\
Moment of inertia of container, $I_{c} = 1$ \\
Number of pinning sites, $N_{\mathrm{pin}} = 10^{5}$ \\
Maximum pinning velocity, $v_{\text{crit}} = 30, 60, \text{and } 90$ \\
Influence of pinning site, $\xi = 0.5\,a$ \\
Runtime of simulation, $T_{\mathrm{run}} = 2000\,T_0$ \\
Initial rotation rate, $\Omega_0 = 50$ \\
Initial rotation period, $T_0 = 2\pi/\Omega_0 = 0.126$ \\
Spin-down rate, $N_{\mathrm{ext}}/I_{c} = -2.5 \times 10^{-4}\,\Omega_0/T_0$ \\
Default integration time-step, $\Delta t = 0.005\,T_0$ \\
\end{tabular}
\end{ruledtabular}
\end{table}

\subsection{Visualization}

We supply three videos, visualizing the first rearrangement in each of the simulations, with the relevant critical unpinning speeds mentioned in the panel headers. The animations, split into two panels, show the evolution of the vortex array on the left, and the velocity field across the superfluid on the right. The black dots in the former are the vortices, where the pinning sites have been made invisible to preserve visual appeal. The arrows in the latter indicate the direction of the local flow, the magnitude given by the colour of the arrow. The vertical colour bar maps the speed of the flow (normalized by $v_{\text{crit}}$) to a gradient ranging from white (weakest), through blue (moderate), to black (strongest). It must be stressed that the length of the arrows, being the same throughout, does not convey the magnitude. Time is tracked at the bottom left of the panels, and the instantaneous rotation rate of the container ($\Omega_{c}$) at the bottom right. The adjacent thin arrows are proxies for the rate change of rotation, with down and up indicating a spinning-down and spinning-up container, respectively. An illustrative snapshot is shown below.

\onecolumngrid
\vspace{5pt}
\begin{center}
    \includegraphics[width=\textwidth]{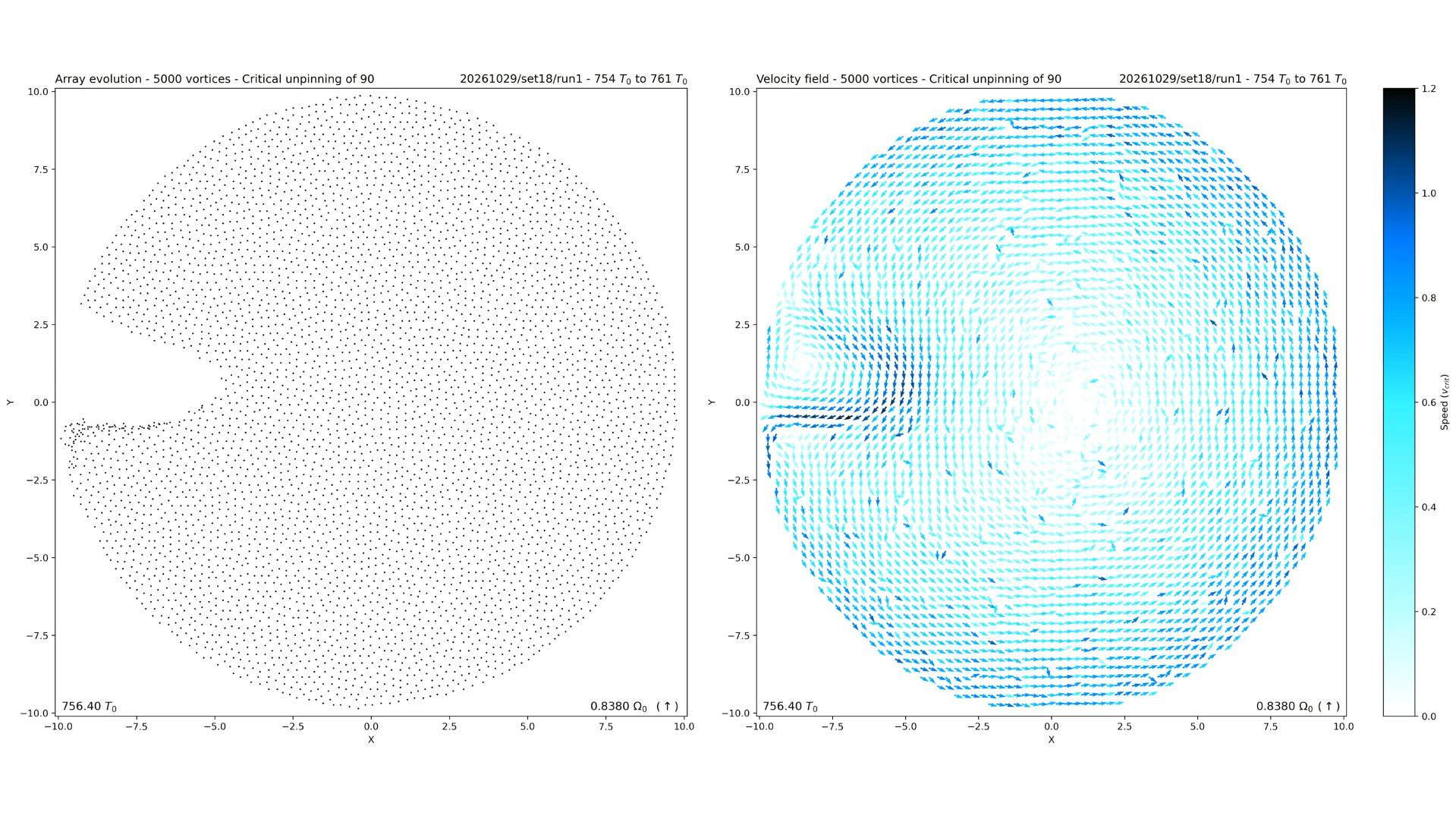}
\end{center}

\end{document}